\RequirePackage{fix-cm}
\documentclass[smallextended]{svjour3}       
\smartqed  
\usepackage{graphicx}
\usepackage[numbers,sort&compress]{natbib}
\usepackage{mathptmx}      
\usepackage{amssymb}
\usepackage{latexsym}
\usepackage[utf8]{inputenc}
\usepackage{nomencl}
\usepackage{natbib}
\usepackage{hyperref}
\usepackage{mathtools}

\newcommand{\ead}{URL: }
\journalname{Journal on Wireless Personal Communications}
\begin{document}

\title{Performance Analysis of FSO System with Spatial Diversity and Relays for M-QAM over Log-Normal Channel}

\author{Pranav Kumar Jha \and Nitin Kachare\and   K Kalyani \and D. Sriram Kumar 
}
\institute{Pranav Kumar Jha \at
              Department of Electronics and Communication Engineering\\
              National Institute of Technology, Tiruchirappalli \\
              Tel.: +91-846-898-3890\\
              \email{jha\_k.pranav@live.com}  \\
         \ead{orcid.org/0000-0001-8053-988X}
}
\date{Received: date / Accepted: date}

\maketitle
\begin{abstract}
The performance analysis of free space optical communication (FSO) systems using relays and spatial diversity at the transmitter end is presented in this paper. The impact of atmospheric turbulence and attenuation caused by different weather conditions and geometric losses has also been taken into account. The effect of turbulence is modeled over a log-normal probability density function. We present the exact closed form expressions of bit-error rate (BER) for M-ary quadrature amplitude modulation (M-QAM). The FSO system link performance is compared for on-off keying, M-ary pulse amplitude modulation and M-QAM modulation techniques. For relay based free space optical communication systems, M-QAM is proved to be superior than other systems considering the same spectral efficiency for each system. A significant performance enhancement in terms of BER analysis and SNR gains is shown for multi-hop MISO FSO system.

\end{abstract}

\section{Introduction}
\label{sec1}
Free space optical communications (FSO) is a rapidly evolving telecommunications to handle high data rate and it has very large information handling capacity. FSO communication systems are presented as an alternative to the fiber optics technology where full duplex data, voice and video transmission in certain applications can happen wirelessly. Even though, light can be competently inserted into fiber cables to route the light information, there are various applications where only the free space between the transmitter and receiver is available to establish a communication link via a clear line-of-sight (LOS) path between transceiver pairs.

Transmission of signals optically in free space has attracted  and opened the door of opportunities towards the areas that has been unexplored since decades. FSO systems are emerging as the fundamental part of multiple systems and applications such as radio frequency (RF) transmission systems, satellite communications, HDTV transmission, optical fiber back up and long-haul connections. FSO systems are  challenged by the atmospheric turbulence, weather attenuation and geometric losses. Changes in refractive index causes scintillation mainly occurs due to atmospheric turbulence by means of temperature, pressure and wind variations \cite{navidpour2007ber}. Different kind of statistical channel models are used to model the turbulence which suits the experimental results. Negative exponential, gamma-gamma and log-normal channel models are used for strong, strong-moderate and weak turbulence conditions, respectively  \cite{luong2013effect,rodrigues2013evaluation,peppas2012simple}. Performance of FSO systems are also largely affected by fading and attenuation due to turbulence in free space medium and different weather conditions (snow, fog and haze), respectively. Additionally, geometric losses due to laser beam divergence also deteriorate the performance of FSO systems. 
Various types of spatial diversity schemes are used to mitigate atmospheric turbulence effects in FSO \cite{abaza2014diversity,tsiftsis2009optical}. Orthogonal Space Time Block Codes (OSTBCs), Transmit Laser Selection (TLS) and Repetition Codes (RCs) are three main spatial diversity techniques used out of which TLS strikes as being the best with channel state information at the transmitter but increases the system complexity \cite{garcia2009selection,safari2008relay}. 
With moderate system complexity in different channel conditions, RCs outperfroms its counterpart OSTBCs. Apart from spatial diversity, some other techniques  have also been proposed to mitigate turbulence effects such as relay assisted techniques \cite{safari2008relay}, error-correcting codes \cite{abaza2011ber,zhu2002free} and maximum likelihood estimation \cite{zhu2000maximum} out of which relay assisted techniques appears to be prominent as several short links are used instead of a long communication link for efficient data transmission over FSO channels.

In this paper, performance of the FSO link for M-ary quadrature amplitude modulation (M-QAM) under different weather conditions has been analyzed using log-normal channel model considering bit error rate (BER) as the performance metric. The effect of turbulence and path losses due to absorption, scattering and scintillation has also been taken into account for analysis purpose. The BER performance analysis has been done for FSO single-input single-output (SISO) system and further improvement has been shown by employing spatial diversity i.e. multiple-input single output (MISO) at the transmitter end. With the help of relays, significant improvement in the link reliability is also discussed and considering the different weather conditions, attenuations and geometric losses, MISO multi-hop system with decode and forward (DF) relays for intensity modulation and direct detection (IM/DD) systems over log-normal channel for short distance communication links is analyzed. With correlation effects among the transmitters, the performance of the multi-hop MISO system using M-QAM is compared to MISO with OOK and RCs, SISO with on-off keying (OOK) and IM/DD SISO multi-hop system using M-PAM \cite{abaza2015performance} while the spectral efficiency is considered to be the same for all systems. At the receiving end, ML decoding receiver is employed for the reception of QAM signal.
\section{FSO System Description} 
\subsection{SISO FSO System using OOK}
\begin{figure*}
\includegraphics[width=0.75\textwidth]{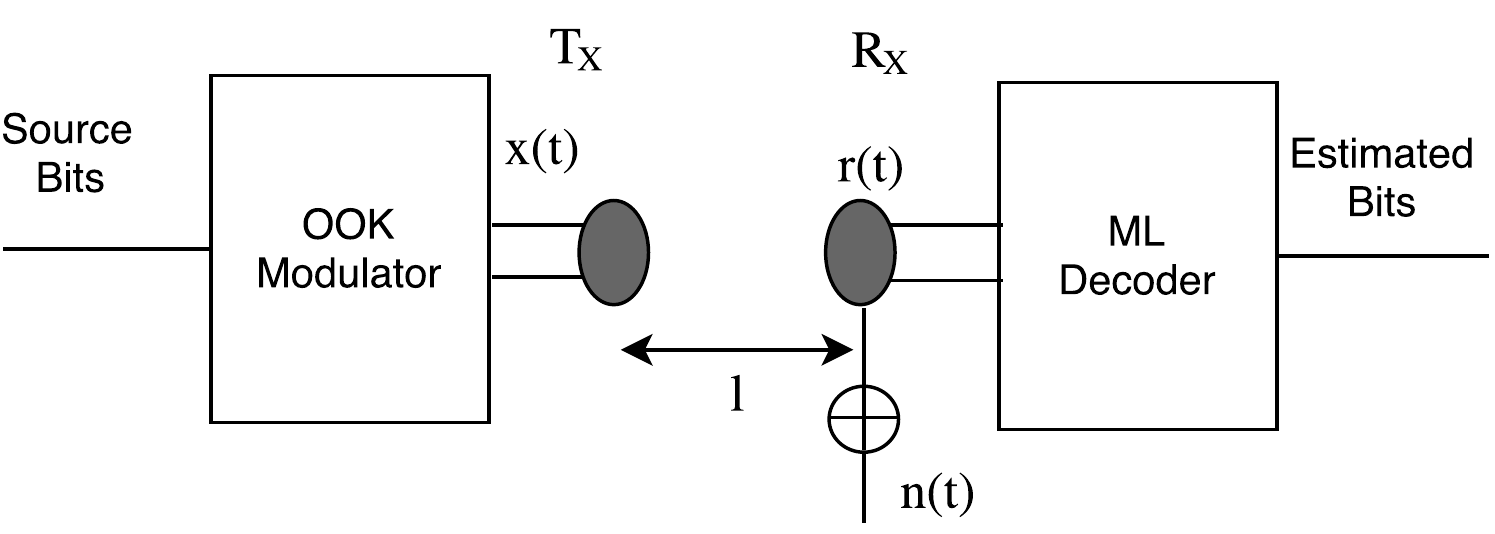}
\caption{SISO FSO system using OOK.}
\label{fig:1}
\end{figure*}
A SISO FSO system is shown in Fig. \ref{fig:1}. Transmission of OOK signals are done using laser and the signal experiences attenuation over a log-normal fading channel because of geometric losses and diverse weather conditions. The signal received at the photodiode is given as
\begin{equation}\label{eq1}
r(t) = x(t)\eta I+n(t),
\end{equation}
where x(t) stands for the transmit signal, $\eta$ is the optical-to-electrical (O/E) conversion coefficient and $I$ implies the received signal beam intensity which is given as \cite{navidpour2007ber}
\begin{equation}\label{eq2}
 I = \beta I_0 h,
\end{equation}
where $I_0$ signifies the received signal beam intensity, $h$ is the channel irradiance exponentially related to an independent and identically distributed (i.i.d.) Gaussian random variable (RV) X having mean $\mu_x$ and variance $\sigma_x^2$ and $\beta$ signifies the normalized path loss coefficient given by $\beta=\frac{\beta(l)}{\beta_d (l)}$, where $\beta_d$ implies the path loss of the direct link and $n(t)$ is an additive white Gaussian noise (AWGN) having mean and variance as zero and $N_{0}/2$, respectively. 

The transmitted beam normally digresses to precise extremity without laser tracking methods which illuminates the receiver which is assumed to be static and spherical wave propagation is considered in this case. Referring to the Rytov theory, the channel variance is given as
\begin{equation}\label{eq3}
 \sigma_x^2 (l) = 0.124 k^{\frac{7}{6}} C_n^2 l^{\frac{11}{6}},
\end{equation}
where $l$ stands for the link distance (km), $k$ implies the wave number and $C_n^2$ signifies the refractive index constant with $\sigma_x (l)\leq0.37$ \cite{abaza2014diversity}. Path loss is calculated as
\begin{equation}\label{eq4}
 \beta(l) = 10^{-\alpha l/10}\frac{D_R^2}{(D_T+\theta_T l)^2},
\end{equation}
where $\alpha$ stands for the attenuation coefficient (dB/km) which is weather dependent, $D_T$ and $D_R$ are the aperture diameters of the transmitters and receivers, respectively and $\theta_T$ signifies the optical light beam divergence angle (mrad).

Dependent on the diferent weather conditions, $\alpha$ varies which makes $h$ as a log-normal RV. The probability density function (pdf) of $h$ is written as
\begin{equation}\label{eq5}
 f(h) = \frac{1}{h\sqrt{8\pi\sigma_x^2}}\textrm{exp}\bigg(\frac{-(\textrm{ln}(h)-2\mu_x)^2}{8\sigma_x^2}\bigg),  \ h \textgreater 0.
\end{equation}

The channel fading coefficients are normalized to unity to save the average power from attenuations or amplification,  which results in $\mu_x=-\sigma_x^2$. By using (\ref{eq2}), at the receiver input, the instantaneous SNR $\gamma$ is given as \cite{tsiftsis2009optical}
\begin{equation}\label{eq6}
 \gamma = \frac{(\eta I)^2}{N_0} =\frac{(\eta \beta I_0 h)^2}{N_0} =\beta^2 \overline{\gamma} h^2,
\end{equation}
where $\overline{\gamma}=\frac{{(\eta I_0)}^2}{N_0}$ stands for the average SNR.
A basic transformation of the RV in (\ref{eq5}) and with the help of (\ref{eq6}), the pdf of the instantaneous SNR is given as
\begin{equation}\label{eq7}
 f(\gamma) = \frac{1}{\sqrt{32 \pi \sigma_x^2}\gamma} \textrm{exp}\Bigg(-\frac{(\textrm{ln}(\frac{\gamma}{\beta^2 \overline\gamma})+4\sigma_x^2)^2}{32\sigma_x^2}\Bigg).
\end{equation}

For any modulation scheme, the approximation of the probability of a symbol error occurred in the observation process is a key criterion and is defined as
\begin{equation*}
  P_{esym}=\textrm{Pr}\{ \hat{ m } \! = m \space| m  \textrm{ sent}\}. 
\end{equation*}
which is difficult to calculate in practice and so, to estimate its value, bouds are used. Using a union bound approximation, a general approximation is given as \cite{Haykin:1988:DC:47310,lee1994digital}
\begin{equation}\label{eq8}
P_{esym} \approx \overline{\textrm{N}}\cdot Q\bigg(\frac{d_{min}}{2\sigma}\bigg),
\end{equation}
where $d_{min}$ represents minimum Euclidean distance between two points in the signal constellation, $\overline{\textrm{N}}$ signifies the average number of constellation points  $d_{min}$ away from each point and
\begin{equation}\label{eq9}
Q(x) \triangleq \frac{1}{2 \sqrt \pi}\int_{u}^{\infty} \textrm{exp}\bigg(\frac{-u^2}{2}\bigg)du,
\end{equation}
implies the standardized Gaussian tail function. For equitably signalling, maximum likelihood (ML) detection is excellent for probability of symbol error and prefers the codeword having the minimum Euclidean distance from the received vector. Due to the exponential characteristics of the $Q(\cdot)$ function, at high SNR ratios, the effect of $d_{min}$  on  $P_{esym}$
is considerably greater than that of  $\overline{\textrm{N}}$ .

In the comparison of different modulation techniques, the probability of bit error, $P_e$ represents a figure of merit which can be resolved from $P_{esym}$ precisely in some cases, while approximations are essential in others. A routine approximation technique is to use Gray coding between contiguous symbols in the constellation. in a Gray coded constellation, adjacent symbols alter by at most a single bit. In consideration of $\textrm{log}_2 M$  bits for every symbol of alphabet size M, the estimation gives \cite{hranilovic2006wireless},
$$P_e\approx\frac{P_{esym}}{\textrm{log}_2 M}.$$

 The conditional bit error probability (BEP) for OOK modulation of a FSO system  is given as \cite{tsiftsis2009optical}
\begin{equation}\label{eq10}
 P(e\textemdash\gamma)\approx Q\bigg(\sqrt{\frac{\gamma}{2}}\bigg),
\end{equation}
where $Q(.)$ represents the Gaussian-$Q$ function. The above equation is simplified with $Q$-function approximation \cite{chiani2003new} on (\ref{eq10}), which yields an approximated closed-form expression as
\begin{equation}\label{eq11}
 P(e\textemdash\gamma) \approx \frac{1}{12} \textrm{exp}\big(-\frac{\gamma}{4}\big)+\frac{1}{4} \textrm{exp}\big(-\frac{\gamma}{3}\big),
\end{equation}

The performance metric i.e. BER can be calculated by evaluating the following integral given by
\begin{equation}\label{eq12}
 \textrm{BER} = \int_0^\infty P(e\textemdash\gamma)f(\gamma)d\gamma.
\end{equation}

By using the Hermite polynomial \cite{abramowitz1964handbook}, the above integral can be simplified  and the approximated closed-form expression can be written as
\begin{equation}\label{eq13}
\begin{multlined}
\textrm{BER} \approx \frac{1}{12 \sqrt{\pi}}\sum_{i=1}^{N} w_i  \textrm{exp}\bigg(-\frac{\beta^2\overline\gamma e^{-4\sigma_x^2+x_i\sqrt{32\sigma_x^2}}}{4}\bigg) 
\\
+ \frac{1}{4\sqrt{\pi}}\sum_{i=1}^{N}w_i\textrm{exp}\bigg(-\frac{\beta^2\overline\gamma e^{-4\sigma_x^2+x_i\sqrt{32\sigma_x^2}}}{3}\bigg),
\end{multlined}
\end{equation}
where $x_i$ implies the zeros and $w_i$ signifies the weights for the order $N$ Hermite polynomial.
\subsection{MISO using RCs System}
\begin{figure*}
\includegraphics[width=0.75\textwidth]{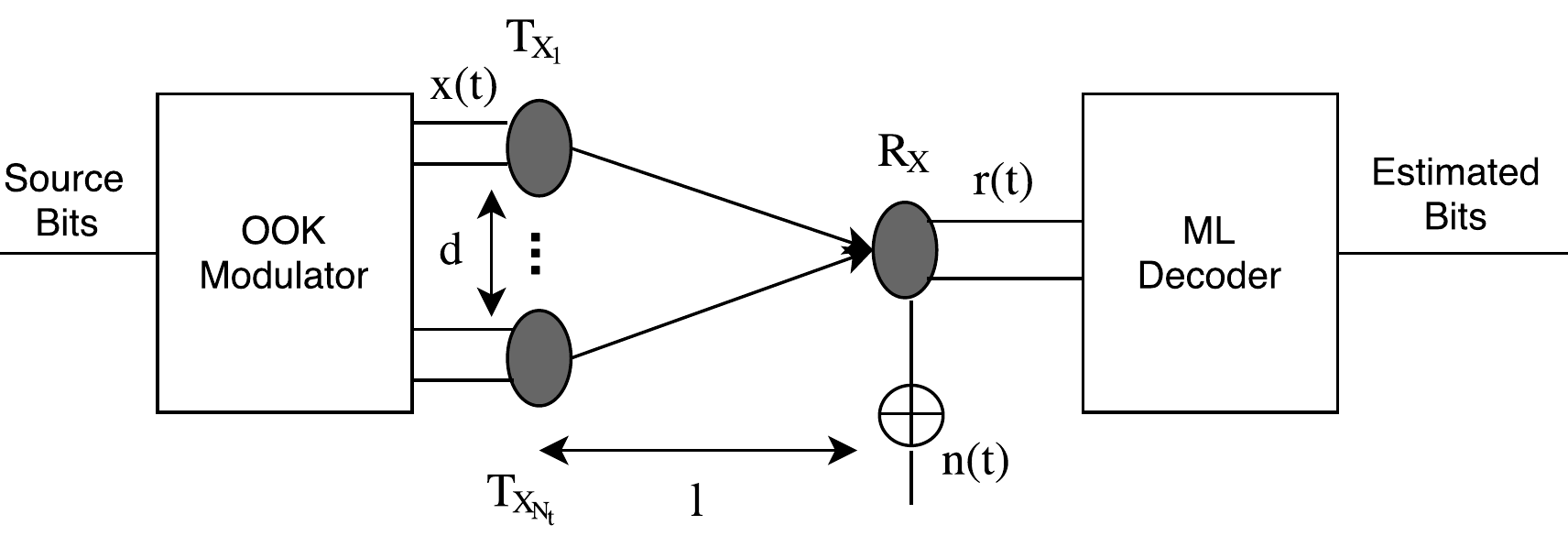}
\caption{MISO FSO system using RCs.}
\label{fig:2}
\end{figure*}
The block diagram for MISO FSO system using RCs with $N_t$ transmitters is shown in Fig. \ref{fig:2}.
In this system, with same statistics as discussed earlier, simultaneous transmissions have been made over correlated log-normal channel for OOK signals through $N_t$ lasers. The  signal received at the photo diode is written as
\begin{equation}\label{eq14}
r(t) = x(t)\eta\sum_{i=1}^{N_t}I_i +n(t).
\end{equation}

Here, $I_i$ signifies the received signal beam intensity from the $i$th transceiver pair. The received signal beam intensity as discussed earlier is written by 
\begin{equation}\label{eq15}
 I_i = \beta I_0 h_i,
\end{equation}
where $h_i = e^{2Y_i}$ is the channel irradiance and $Y_i$ is a correlated and i.i.d. RV with mean $\mu_y$ and variance $\sigma_y^2$ which can be calculated as (\ref{eq3}).

The overall spacing between the transmitters should be smaller than $\theta l$ meters, where $\theta$ signifies the field of view (FOV) of the receiver in radian 
(rad). $\Gamma$ signifies the spatial covariance matrix  which models the spatial correlation among the transmitters written as \cite{navidpour2007ber}
\begin{equation}\label{eq16}
 \Gamma_{ij} = \sigma_y^{2}\times\rho^{\textemdash i-j\textemdash} ,
\end{equation}
where $\rho$ stands for the correlation coefficient, $i$ and $j$ represents the row and column indices of the coefficients of the covariance matrix, respectively and $|.|$ represents the absolute value. $\rho$ is given as a function of separation distance $d$ among transmitters which is written as \cite{zhu2000maximum}
\begin{equation}\label{eq17}
\rho = \textrm{exp}\bigg[-\bigg(\frac{d}{d_0}\bigg)^{\frac{5}{3}}\bigg],
\end{equation}
where the correlation length $d_0\approx\sqrt{\lambda l}$ where $\lambda$ implies the wavelength (nm). 

As $l\gg d$, all transceiver pair apertures are considered to be equidistant. 

Following the steps discussed in (\ref{eq7}), the pdf for the instantaneous SNR can be written as
\begin{equation}\label{eq18}
f(\gamma_1,\gamma_2\ldots \gamma_{N_t}) = \frac{\textrm{exp}\bigg(-\frac{1}{32} Z(\Gamma)^{-1}Z^T\bigg)}{4^{N_t}\sqrt{(2\pi)^{N_t}(\textrm{det}[\Gamma])}\prod_{i=1}^{N_t}\gamma_i},
\end{equation}
where $Z = \bigg[ln\bigg(\frac{\gamma_1}{(\beta^2\overline\gamma_1)}\bigg)\ldots ln\bigg(\frac{\gamma_{N_t}}{\beta^2 \overline\gamma_{N_t}}\bigg)\bigg]-4\mu_y$ and $(\cdot)^{-1}$ and $(\cdot)^T$ represents the inverse and transpose of a matrix $(\cdot)$, respectively. 

The MISO FSO system for OOK using RCs has the conditional BEP given by \cite{abaza2014diversity}
\begin{equation}\label{eq19}
P(e|\gamma) = Q\bigg(\frac{1}{N_t}\sum_{i=1}^{N_t}\sqrt{\frac{\gamma_i}{2}}\bigg),
\end{equation}
where the $\frac{1}{N_t}$ term represents the same power as conventional SISO FSO system. 

The BER for the FSO MISO system is written as \cite{abaza2015performance}
\begin{equation}\label{eq20}
\begin{multlined}
\textrm{BER}\approx
\int_0^\infty\ldots \int_0^\infty\frac{1}{12}\textrm{exp}\bigg\{-\frac{1}{2}\bigg(\frac{1}{N_t}\sum_{i=1}^{N_t}\sqrt{\frac{\gamma_i}{2}}\bigg)^2\bigg\}\\
\times f(\gamma_1,\gamma_2\ldots \gamma_{N_t}) d\gamma_1\ldots d\gamma_{N_t}\\
+ \int_0^\infty\ldots\int_0^\infty\frac{1}{4}\textrm{exp}\bigg\{-\frac{2}{3}\bigg(\frac{1}{N_t}\sum_{i=1}^{N_t}\sqrt{\frac{\gamma_i}{2}}\bigg)^2\bigg\}\\
\times f(\gamma_1,\gamma_2\ldots \gamma_{N_t})d\gamma_1\ldots d\gamma_{N_t}.
\end{multlined}
\end{equation}

The average BER can be calculated using moment generating function (MGF) as \cite{abaza2014diversity}
\begin{equation}\label{eq21}
\begin{multlined}
\textrm{BER} \approx \sum_{n_1=1}^{N}\ldots \sum_{n_{N_t}=1}^{N}\bigg[\prod_{i=1}^{N_t}\frac{w_{n_i}}{\sqrt{\pi}}\bigg]
\\
\times\frac{1}{12}\textrm{exp}\bigg(-\frac{\beta^2\overline\gamma}{4N_t}\sum_{i=1}^{N_t}\bigg[\textrm{exp}\bigg(\sqrt{32}\sum_{j=1}^{N_t}c_{ij}x_{nj}-4\sigma_y^2\bigg)\bigg]\bigg)
\\
+\sum_{n_1=1}^{N}\ldots \sum_{n_{N_t}=1}^{N}\bigg[\prod_{i=1}^{N_t}\frac{w_{n_i}}{\sqrt{\pi}}\bigg]
\\
\times\frac{1}{4}\textrm{exp}\bigg(-\frac{\beta^2\overline\gamma}{3N_t}
\sum_{i=1}^{N_t}\bigg[\textrm{exp}\bigg(\sqrt{32}\sum_{j=1}^{N_t}c_{ij}x_{nj}-4\sigma_y^2\bigg)\bigg]\bigg),
\end{multlined}
\end{equation}
with $x_{n_j}$ signifies the zeros and $w_{n_i}$ implies the weights of the order $N$ Hermite polynomial and $c_{ij}$ signifies the $(i,j)$th coefficient of $\Gamma_{sq}=\Gamma^{1/2}$.
\subsection{Multi-hop DF Relaying System}
\subsubsection{M-QAM Transmission}
\begin{figure*}
\includegraphics[width=0.75\textwidth]{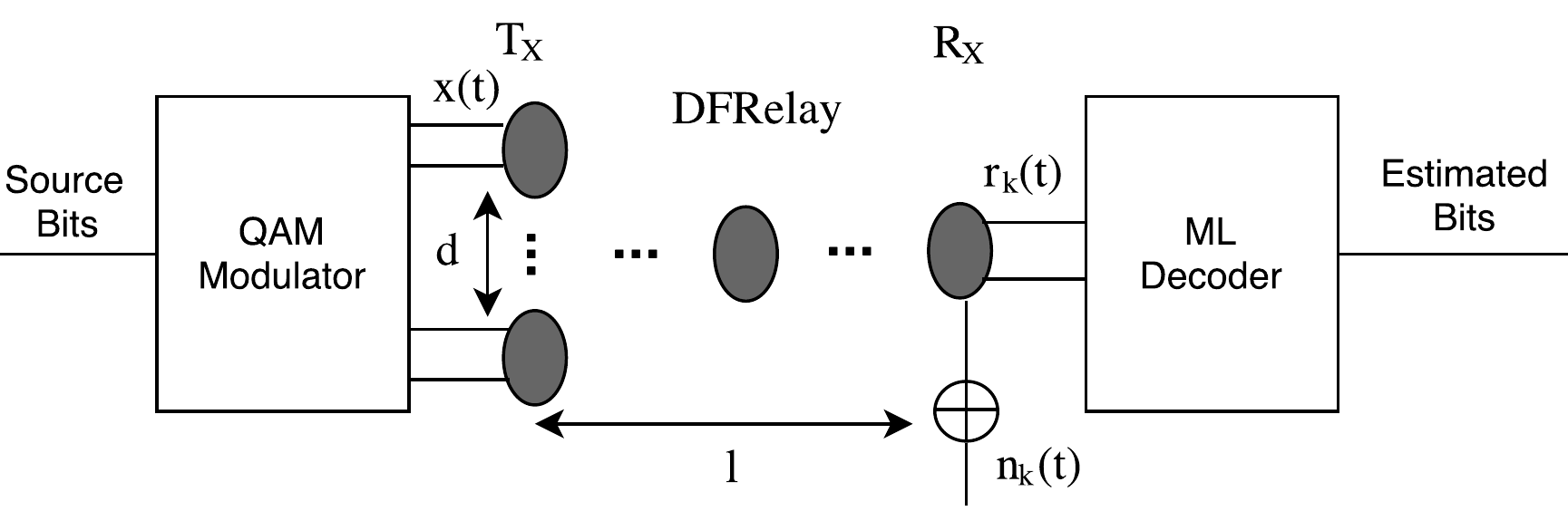}
\caption{MISO FSO system using Relays.}
\label{fig:3}
\end{figure*}
A MISO FSO system using M lasers and a photo detector is shown in Fig. \ref{fig:3}. At the transmitter end, the binary data is modulated using an electrical $M_I \times M_Q$ QAM modulator and the Gray coding is used for mapping data into QAM symbols and then transmitted using all the M paths. Additionally the channel distance is also assumed to be sufficient in order to neglect the effect of spatial correlation. At the receiver, ML decoding technique is used for the combination of optical signal.

Rectangular QAM signal constellations can be easily generated as two PAM signals using phase-quadrature carriers and demodulation is easy for such a signal. For $M \geq 16$, despite of not being the best  M-ary QAM constellations, the average transmit power needed to acquire a stated minimum distance is uniquely marginally higher than the average power needed for the perfect M-ary QAM constellation. These are the logics for which rectangular M-ary QAM signals are repeatedly used QAM signals. For rectangular signal constellations where $M=2^k$ with $k$ even, the QAM signal constellation is equivalent to two PAM signals on quadrature carriers, all having $\sqrt M= 2^{k/2}$ points. The probability of error of QAM can be efficiently calculated from the probability of error of PAM since the signals in the phase-quadrature components can be perfectly abstracted at the QAM demodulator,. Specifically, the probability of a correct decision for the M-ary QAM system can be gives as \cite{proakis2001digital}
\begin{equation}\label{eq22}
P_C= (1-P_{\sqrt M} )^2,
\end{equation}
where $P_{\sqrt M}$ represents the probability of error for $\sqrt {M}$-ary PAM having half the average power in every quadrature signal of the correspondent QAM system. An appropriate modification of the probability of error of M-ary PAM yields
\begin{equation}\label{eq23}
P_{\sqrt M}=2\bigg(1-\frac{1}{\sqrt M} \bigg)Q\bigg(\frac{3log_2 M\gamma}{M-1}\bigg),
\end{equation}
where $\gamma$ stands for the average SNR per symbol. Therefore for the M-ary QAM, the probability of a symbol error can be written as
\begin{equation}\label{eq24}
P_M=1-(1-P_{\sqrt M})^2.
\end{equation}
It is valid for $M=2^k$ with $k$ even. There is no correspondent $\sqrt M$-ary PAM system for odd $k$ and the error rate of a rectangular signal set can be determined comfortably. By employing the optimum detector, the symbol error probability must be tightly upper bounded as
\begin{equation}\label{eq25}
\begin{multlined}
P_M\leq 1-\bigg[1-2Q\bigg(\sqrt{\frac{3 \gamma}{M-1}}\bigg)\bigg]\\
\leq 4Q\bigg(\sqrt{\frac{3 log_2{M}\gamma}{M-1}}\bigg).
\end{multlined}
\end{equation}

Here, the effect of introducing relays in FSO system for SISO and MISO configurations has been considered using DF relays and significant improvent in the link performance is analyzed. $(K-1)$ DF relays are positioned for the transceiver pair for $K$ hops and the signal received at the $k$th hop can be written as 
\begin{equation}\label{eq26}
r_k(t) = x(t)\eta I_k + n_k(t), \ k=1,2,\ldots K.
\end{equation}

In this technique, unlike SISO and MISO systems where one-time slot is required,  $K$ time slots are required for transmission of the signal from source to destination. To achieve the similar spectral efficiency, $2^K$-ary M-PAM \cite{abaza2015performance} and M-QAM modulation techniques are considered. The effects of turbulence and path losses are significantly reduced by creating shorter communication links using DF relays.
For multi-hop DF relay system, the upper bound BER is written as
\begin{equation}\label{eq27}
\textrm{BER} \leq 1-\prod_{k=1}^{K}(1-\textrm{BER}_k).
\end{equation}

Considering the identical statistical properties for all hops, above equation can be approximated by  
\begin{equation}\label{eq28}
\textrm{BER} \approx \frac{1}{2}[1-(1-2\textrm{BER}_k )^K].
\end{equation}

The conditional BEP of M-QAM is given by \cite{singh4performance} as
\begin{equation}\label{eq29}
P(e\textemdash\gamma)\approx \frac{2\Big(1-\frac{1}{\sqrt M}\Big)}{\textrm{log}_2 M}\Bigg[Q\Bigg(\sqrt{\frac{3\textrm{log}_2M \gamma}{2(M-1)}} \Bigg)\Bigg].
\end{equation}

The instantaneous SNR is given by 
\begin{equation}\label{eq30}
\gamma = \frac{2P^2}{\sigma_n^2 R},
\end{equation}
where $P$ signifies the signal power and $R$ implies the bit rate. 
Now, applying the approximate $Q$ function on (\ref{eq29}), the conditional BEP of M-QAM is
\begin{equation}\label{eq31}
\begin{multlined}
P(e\textemdash\gamma)\approx \frac{2\Big(1-\frac{1}{\sqrt M}\Big)}{\textrm{log}_2 M}
\times\bigg[\frac{1}{12}\textrm{exp}\bigg(-\frac{3 \textrm{log}_2 M \gamma}{4(M-1)}\bigg)
+ \frac{1}{4}\textrm{exp}\bigg(-\frac{ \textrm{log}_2 M \gamma}{(M-1)}\bigg)\bigg],
\end{multlined}
\end{equation}
which represents the typical expression for the conditional BEP of the FSO system with M-QAM. BER expression for the $k$th hop can be derived by using the steps discussed earlier and with the definition of Hermite polynomial \cite{abaza2015performance} and $Q$-function approximation. The closed form expression of BER for M-QAM modulation is derived and given by
\begin{equation}\label{eq32}
\begin{multlined}
\textrm{BER} \approx \frac{G}{12}\sum_{i=1}^{N}w_i \textrm{exp}\bigg(-\frac{3\textrm{log}_2M \beta_{kn}^2 \overline\gamma e^{-4\sigma_k^2+x_i\sqrt{32\sigma_k^2}}}{4(M-1)}\bigg)
\\
+\frac{G}{4}\sum_{i=1}^{N}w_i \textrm{exp}\bigg(-\frac{\textrm{log}_2M \beta_{kn}^2 \overline\gamma e^{-4\sigma_k^2+x_i\sqrt{32\sigma_k^2}}}{(M-1)}\bigg),
\end{multlined}
\end{equation}
where $G = \frac{2\big(1-\frac{1}{\sqrt M}\big)}{\textrm{log}_2 (M) \sqrt{\pi}}$ and $\beta_{kn}$ is the normalized path loss coefficient with reference to the direct link for the multi-hop system. Substituting (\ref{eq32}) in (\ref{eq27}) and (\ref{eq28}), the upper bound BER and the average BER  expressions for the multi-hop system are obtained, respectively. The BER expression for M-PAM is given as Eq. (26) in \cite{abaza2015performance}.
\subsubsection{$M^2$-QAM Transmission}
The  $M^2$ symbols of $M^2$-QAM  made up of an in-phase and quadrature phase component basis function, orthogonal to each other. In each symbol duration, the two basis functions is modulated with the independent data resulting a multiplication by a series of M amplitude values to each basis function to comprise the $M^2$ symbols \cite{hranilovic2006wireless}. 

The constellation of QAM shows a two dimensional regular array of points and the minimum spacing between the points is prescribed by the amount of DC bias appended by virtue of the non-negativity constraint which is \cite{hranilovic2006wireless}
\begin{equation*}
d_{min}=\frac{P}{M-1}\sqrt{\frac{2\textrm{log}_2 M}{R}},
\end{equation*}
where R signifies the bit rate.

The probability of symbol error is estimated by using the union bound approximation in (\ref{eq9}). By evaluating the average number of neighbours $d_{min}$ away from every constellation point, $P_{esym}$ is approximated as
\begin{equation*}
P_{esym}=\frac{4M-1}{m}\cdot Q\Bigg(\frac{P}{M-1}\sqrt{\frac{1}{4 R_s \sigma^2}}\Bigg),
\end{equation*}
where $R_s=R/\textrm{log}_2 M^2$.

Applying the Gray coding approximation, the conditional BEP of $M^2$-QAM is given by \cite{hranilovic2006wireless}
\begin{equation}\label{eq33}
P(e\textemdash\gamma)\approx \frac{2\Big(M-1\Big)}{M\textrm{log}_2M}\Bigg[Q\Bigg(\sqrt{\frac{\textrm{log}_2M \gamma}{4{(M-1)}^2}}\Bigg)\Bigg].
\end{equation}

Where, instantaneous SNR 
$
\gamma = \frac{2P^2}{\sigma_n^2 R}.
$
Now, applying the approximate $Q$ function on (\ref{eq33}), the conditional BEP of M-QAM is given as
\begin{equation}\label{eq34}
P(e\textemdash\gamma)\approx \frac{2\Big(M-1\Big)}{M\textrm{log}_2M}
\times\bigg[\frac{1}{12}\textrm{exp}\bigg(-\frac{ \textrm{log}_2 M \gamma}{8(M-1)^2}\bigg)
+ \frac{1}{4}\textrm{exp}\bigg(-\frac{\textrm{log}_2 M \gamma}{6(M-1)^2}\bigg)\bigg],
\end{equation}
which represents the typical expression for the conditional BEP of the FSO system with $M^2$-QAM. The closed form expression of BER for $M^2$-QAM modulation derived as discussed earlier, can be written as
\begin{equation}\label{eq35}
\begin{multlined}
\textrm{BER} \approx \frac{G}{12}\sum_{i=1}^{N}w_i \textrm{exp}\bigg(-\frac{\textrm{log}_2M \beta_{kn}^2 \overline\gamma e^{-4\sigma_k^2+x_i\sqrt{32\sigma_k^2}}}{8(M-1)^2}\bigg)
\\
+\frac{G}{4}\sum_{i=1}^{N}w_i \textrm{exp}\bigg(-\frac{\textrm{log}_2M \beta_{kn}^2 \overline\gamma e^{-4\sigma_k^2+x_i\sqrt{32\sigma_k^2}}}{6(M-1)^2}\bigg),
\end{multlined}
\end{equation}
where $G = \frac{2\big(M-1\big)}{M\textrm{log}_2M \sqrt{\pi}}$ and $\beta_{kn}$ is the normalized path loss coefficient with reference to the direct link for the multi-hop system . Substituting (\ref{eq35}) in (\ref{eq27}) and (\ref{eq28}), the upper bound BER expression and the average BER  expression for the multi-hop system are retrieved, respectively.
\subsection{Multi-hop MISO DF Relaying System}
\begin{figure*}
\includegraphics[width=0.75\textwidth]{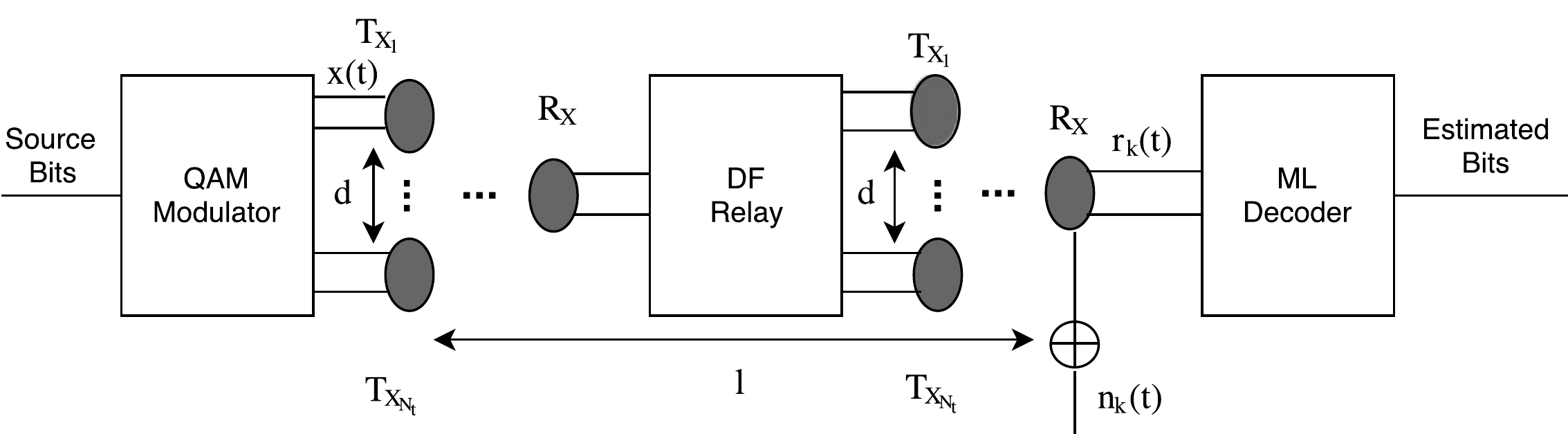}
\caption{Multi-hop MISO system with DF Relays.}
\label{fig:4}
\end{figure*}

This is a hybrid scheme which utilizes the superiority of both the systems to lessen the effect of channel attenuations and turbulence as the effect of relays as well as transmit diversity has been considered simultaneously. The block diagram for this system is shown in Fig. \ref{fig:4}.

There are $(K-1)$ relays which receives the signal from $N_t$ transmitters and retransmits it simultaneously using RCs. In this system model, the number of transmitters per relay is considered to be identical with hops. The signal received at every hop can be written as
\begin{equation}\label{eq36}
r_k (t) = x(t)\eta\sum_{i=1}^{N_t}I_{ki} +n_k (t),  \ k=1,2 \ldots K.
\end{equation}

Following the similar approach and using conditional BEP equation of M-QAM, the BER expression for the $k$th hop has been derived and is given by
\begin{equation}\label{eq37}
\begin{multlined}
\textrm{BER}_k \approx
\sum_{n_1=1}^{N}\ldots \sum_{n_{N_t}}^{N}\Bigg[\prod_{i=1}^{N_t}\frac{w_{n_i}}{\sqrt{\pi}}\Bigg]
\\\times
\frac{F}{12}\textrm{exp}\Bigg(-\frac{3\textrm{log}_2(M) \beta_{kn}^2\overline\gamma}{4(M-1)N_t}
\sum_{i=1}^{N_t}\Bigg[\textrm{exp}\bigg(\sqrt{32}\sum_{j=1}^{N_t}{c_{ij}^{'}x_{nj}-4\sigma_k^2}\bigg)\Bigg]\Bigg)\\
+\sum_{n_1=1}^{N}\ldots \sum_{n_{N_t}}^{N}\Bigg[\prod_{i=1}^{N_t}\frac{w_{n_i}}{\sqrt{\pi}}\Bigg]
\\\times
\frac{F}{4}\textrm{exp}\Bigg(-\frac{\textrm{log}_2(M) \beta_{kn}^2\overline\gamma}{(M-1)N_t}
\sum_{i=1}^{N_t}\Bigg[\textrm{exp}\bigg(\sqrt{32}\sum_{j=1}^{N_t}{c_{ij}^{'}x_{nj}-4\sigma_k^2}\bigg)\Bigg]\Bigg).
\end{multlined}
\end{equation}

And, using the conditional BEP equation of $M^2$-QAM, the BER expression for the $k$th hop has been given as
\begin{equation}\label{eq38}
\begin{multlined}
\textrm{BER}_k \approx
\sum_{n_1=1}^{N}\ldots \sum_{n_{N_t}}^{N}\Bigg[\prod_{i=1}^{N_t}\frac{w_{n_i}}{\sqrt{\pi}}\Bigg]
\\\times
\frac{F}{12}\textrm{exp}\Bigg(-\frac{\textrm{log}_2(M) \beta_{kn}^2\overline\gamma}{8(M-1)N_t}
\sum_{i=1}^{N_t}\Bigg[\textrm{exp}\bigg(\sqrt{32}\sum_{j=1}^{N_t}{c_{ij}^{'}x_{nj}-4\sigma_k^2}\bigg)\Bigg]\Bigg)\\
+\sum_{n_1=1}^{N}\ldots \sum_{n_{N_t}}^{N}\Bigg[\prod_{i=1}^{N_t}\frac{w_{n_i}}{\sqrt{\pi}}\Bigg]
\\\times
\frac{F}{4}\textrm{exp}\Bigg(-\frac{\textrm{log}_2(M) \beta_{kn}^2\overline\gamma}{6(M-1)N_t}
\sum_{i=1}^{N_t}\Bigg[\textrm{exp}\bigg(\sqrt{32}\sum_{j=1}^{N_t}{c_{ij}^{'}x_{nj}-4\sigma_k^2}\bigg)\Bigg]\Bigg),
\end{multlined}
\end{equation}
where $F = \frac{2\big(1-\frac{1}{\sqrt M}\big)}{\textrm{log}_2 (M)}$ for (\ref{eq37})  and $F=\frac{2\big(M-1\big)}{M\textrm{log}_2M}$ for (\ref{eq38}) and $c_{ij}^{'}$ denotes the $(i,j)$th coefficients of  $\Gamma_{sq}^{'}=\Gamma^{'1/2}$. These results can be substituted in (\ref{eq27}) and (\ref{eq28}) to get the expressions of the upper bound BER  and the final approximated mathematical closed form explanation for the average BER of the multi-hop MISO system, respectively. The BER expression for M-PAM multi-hop FSO system is given as Eq. (28) in \cite{abaza2015performance}. 
\begin{table}[htbp]
\begin{center}
\caption{Simulation Parameters}
\label{tab:1} 
\begin{tabular}{ll}
\hline\noalign{\smallskip}
\textbf{FSO Parameters} & \textbf{Numerical Values }  \\
\noalign{\smallskip}\hline\noalign{\smallskip}
Relay spacing & 400 m \\
Wavelength ($\lambda$) & 1550 nm  \\
Link distance $(l)$ & 1200 m\\
Beam divergence\\ angle $(\theta_T)$ &	2 mrad\\
Correlation coefficient $(\rho)$ & 0.3\\
Transmitter and receiver  \\aperture diameter\\ $(D_R\textrm{ and }D_T)$	&
20 cm  \\
Attenuation\\ constant $(\alpha)$ & 0.43 dB/km (clear weather)\\&
20 dB/km (light fog)\\
Refractive index\\ constant $(C_n^2)$ &
$5\times10^{-14} \textrm{m}^{-(2/3)}$ (clear weather)\\&
$1.7\times 10^{-14} \textrm{m}^{-2/3}$
(light fog)\\
\noalign{\smallskip}\hline
\end{tabular}
\end{center}
\end{table}

\section{Numerical Analysis and Discussions}
\begin{figure*}
\includegraphics[width=0.75\textwidth]{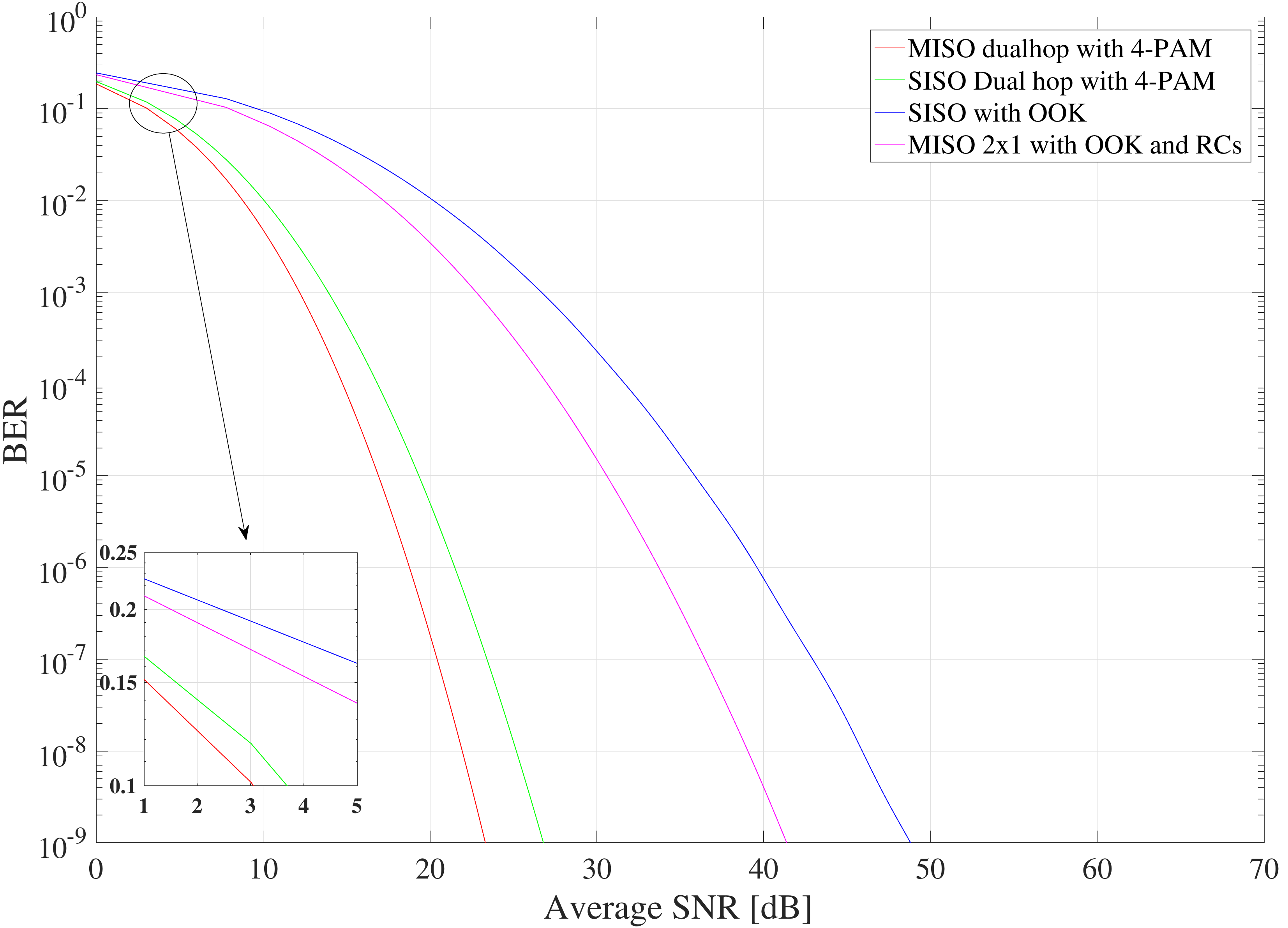}
\caption{BER for dual-hop FSO system with 4-PAM, SISO with OOK and MISO with OOK and RCs under clear weather conditions.}
\label{fig:5}
\end{figure*}

\begin{figure*}
\includegraphics[width=0.75\textwidth]{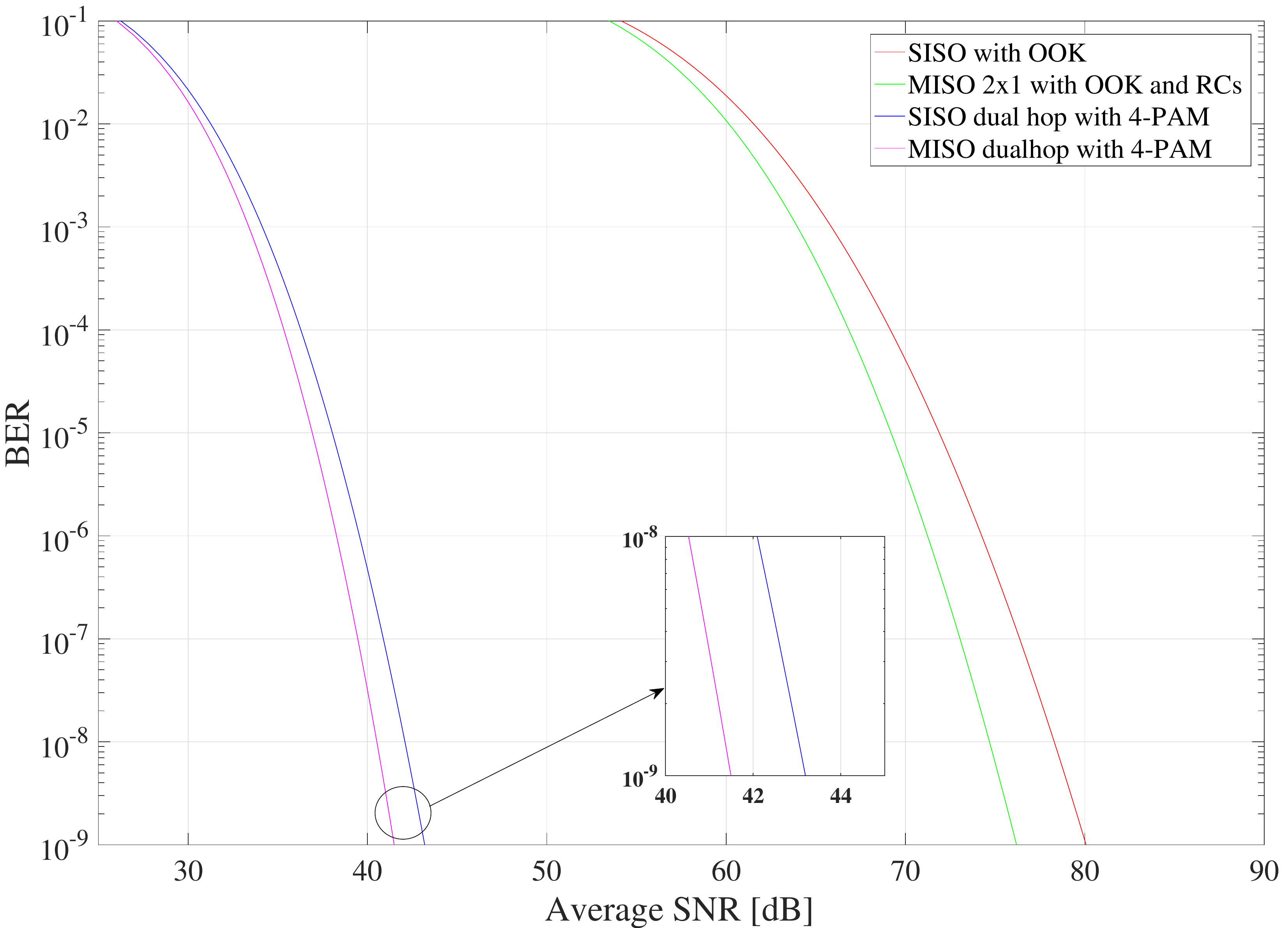}
\caption{BER for dual-hop FSO system with 4-PAM, SISO with OOK and MISO with OOK and RCs under light fog conditions.}
\label{fig:6}
\end{figure*}

For evaluation purpose, numerical results are furnished to check the consistency of the derived analysis presented in this paper. All the parameters considered to obtain the results are tabulated in Table \ref{tab:1} where the objective is to acheive a target BER of $10^{-9}$, $\lambda=1550$ nm, $D_R=D_T=20$ cm, $\theta_T=2$ mrad, $l = 1200$ m, $C_n^2= 1.7 \times 10^{-14} \textrm{m}^{-2/3}$ and $\alpha=20$ dB/km for light fog condition in the morning,  $C_n^2= 5 \times 10^{-14} \textrm{m}^{-2/3}$ and $\alpha=20$ dB/km for clear
weather, $\rho$ = 0.3 and
all hops are equidistant. For
simulation purpose, a minimum of $10^6$ bits are relayed for respective SNR
values and increases with increasing SNR. The scintillation index (SI) is $\leq$ 0.75 for log-normal
channel and 
$\sigma_x=\sqrt{\textrm{ln}(\textrm{SI})+1/2}$ \cite{moradi2011switched}. Hence, for SI = 0.75, $\sigma_x\leq0.374$ is required, which represents the maximal value assumed for the analysis purpose. Fig. \ref{fig:5} and Fig. \ref{fig:6} are regenerated from the results of \cite{abaza2015performance} for analysis and comparison purpose and shows SISO with OOK and MISO with OOK and RCs under clear weather and light fog conditions, respectively. 
\begin{figure*}
\includegraphics[width=0.75\textwidth]{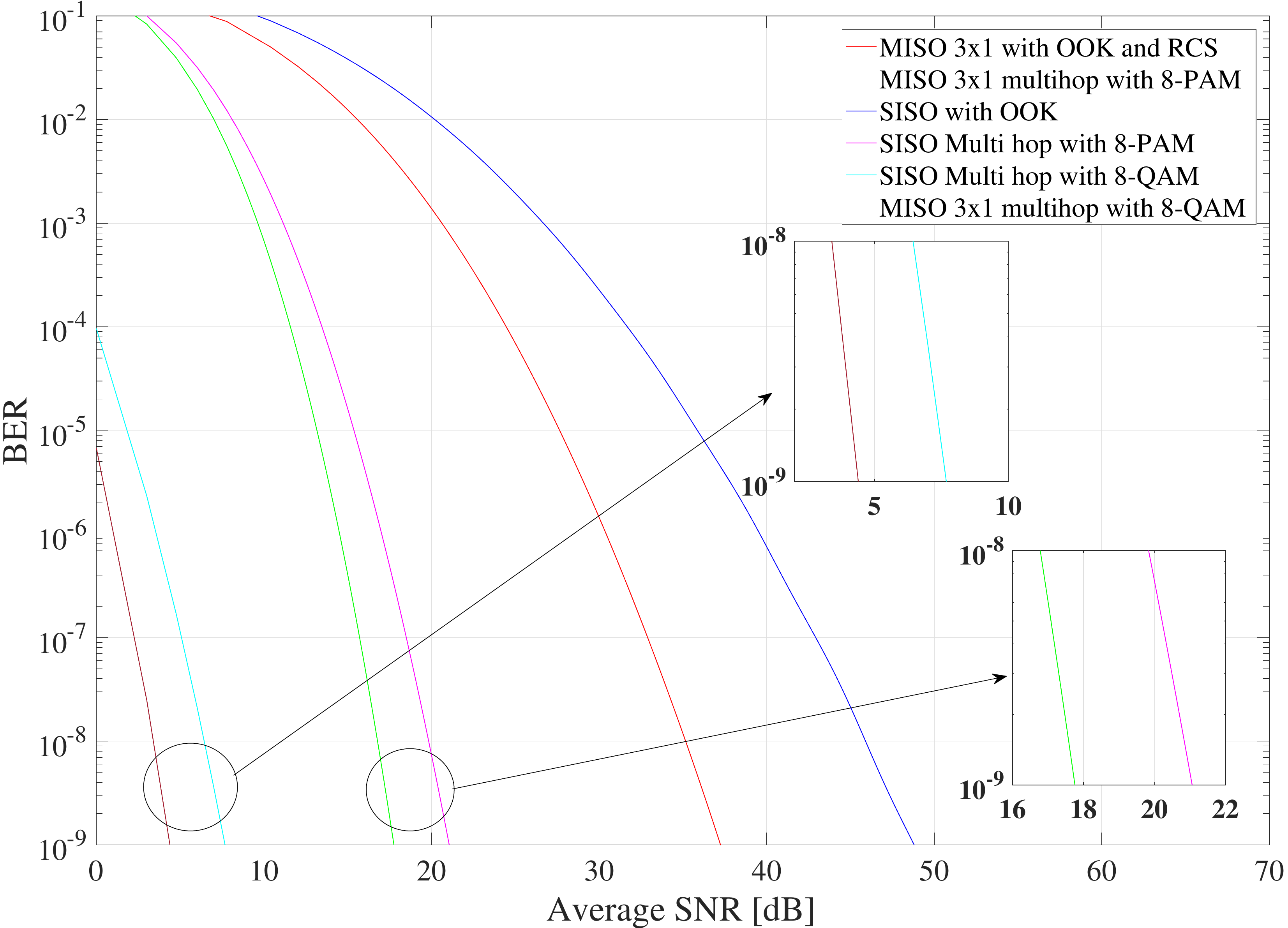}
\caption{BER for two-relay multi-hop FSO system  with 8-QAM, 8-PAM, SISO with OOK and MISO with OOK and RCs under clear weather conditions.}
\label{fig:7}
\end{figure*}
\begin{figure*}
\includegraphics[width=0.75\textwidth]{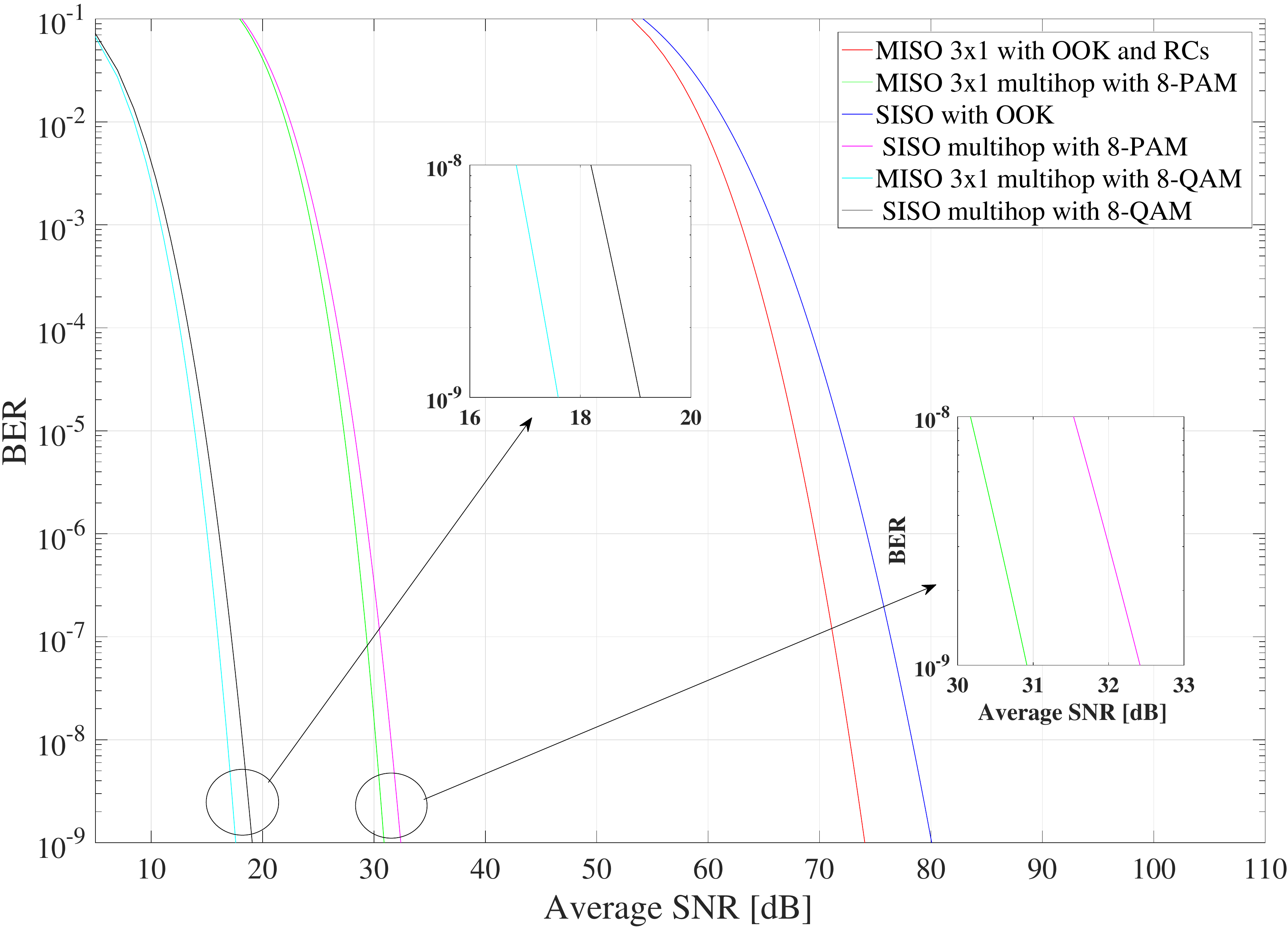}
\caption{BER for two-relay multi-hop FSO system with 8-QAM, 8-PAM, SISO with OOK and MISO with OOK and RCs under light fog conditions.}
\label{fig:8}
\end{figure*}

Fig. \ref{fig:5} represents the BER for dual-hop FSO system by applying either one or two transmitters using 4-PAM modulation, MISO with OOK and RCs using two transmitters and SISO with OOK modulation and shows the performance improvement for the multi-hop SISO system for clear weather conditions with moderate turbulence where a relay is placed at the middle of the source and destination and 4-PAM modulation technique is used for the same spectral efficiency as MISO and SISO systems. The multi-hop SISO and multi-hop MISO systems using PAM have 18 dB and 22 dB SNR gains at the target BER in comparison with the SISO with OOK and MISO with OOK and RCs systems. The multi-hop MISO system surpasses the multi-hop SISO system by 3.4 dB. 

Fig. \ref{fig:6} represents BER for dual-hop FSO system by applying either one or two transmitters using 4-PAM modulation scheme, MISO with OOK and RCs using two transmitters and SISO with OOK modulation scheme and shows the effect of light fog weather conditions on the performance of every system where low turbulence is considered as fog can not occur in vigorous sunlight. The Performance of the systems has been degraded by light fog. SISO  and MISO systems are strongly influenced by light fog compared to multi-hop systems. At the target BER, loss of 31 dB and 35 dB has been calculated for SISO and MISO systems respectively where in case of multi-hop systems, it is not more than 18 dB. The multi-hop SISO  system is still superior to other systems and in comparison with SISO and MISO systems, SNR gains of about 37 dB and 33 dB can be acquired, respectively.

Fig. \ref{fig:7} represents BER for two relay multi-hop FSO system by applying either one or three transmitters using 8-QAM, 8-PAM modulation, MISO with OOK and RCs with three transmitters and SISO with OOK modulation and shows the effect of increasing the transmit diversity to three in a MISO system and the sum of relays upto two in a multi-hop system for clear weather conditions. At the target BER, SNR gains of 26 dB and 19.5 dB is achieved at the target BER for multi-hop SISO and multi-hop MISO systems with three transmitters as compared to SISO with OOK and MISO with OOK and RCs using three transmitters for 8-PAM modulation whereas, using 8-QAM modulation technique, SNR gains of 14 dB and 13 dB is achieved for multi-hop SISO and multi-hop MISO with three transmitters as compared to multi-hop SISO and multi-hop MISO using three  transmitters with PAM. Multi-hop MISO system surpasses the multi-hop SISO system by 2.8 dB and 2.6 dB in case of M-PAM and M-QAM, respectively.

Fig. \ref{fig:8} represents BER for two relay multi-hop FSO system by applying either one or three transmitters using 8-QAM, 8-PAM modulation, MISO with OOK and RCs with three transmitters and SISO with OOK modulation under light fog conditions. For two relay multi-hop system, 8-QAM modulation provides an equal spectral efficiency related to SISO and MISO systems which is in parallel with the consideration of the same spectral efficiency for all systems. Analytic calculations demonstrate that systems having multiple relays or multi-hop FSO systems are still dominating over other systems and the performance of the systems has increased significantly by 10 dB. With the increasing number of transmitters to three, a increment of 2 dB for M-PAM and 1.8 dB for M-QAM has been shown at the target BER. SNR gains of 47.5 dB and 42.8 dB is achieved at the target BER for multi-hop SISO  and multi-hop MISO as compared to SISO with OOK and MISO with OOK and RCs for 8-PAM modulation whereas, using 8-QAM modulation technique, SNR gains of 12.5 dB and 13 dB is achieved for multi-hop SISO and multi-hop MISO with three transmitters as compared to multi-hop SISO and multi-hop MISO using single transmitter with PAM. Multi-hop MISO system surpasses the multi-hop SISO system by 1.4 dB and 1.2 dB in case of M-PAM and M-QAM, respectively.
\section{Conclusions}
The effect of diverse atmospheric acclimations on the  multi-hop MISO DF relay FSO system has been analyzed and different path loss attenuations have also been included for analysis purpose. Analysis shows that MISO multi-hop system using M-QAM modulation technique has significant improvement over other systems such as SISO with OOK and MISO with OOK and RCs, SISO multi-hop system using M-PAM having the equal spectral efficiency. The effects of clear weather conditions with moderate turbulence and light fog with weak turbulence on the the behaviour of different configurations of FSO system are investigated. Multi-hop systems can counteract the effects of both turbulence and path losses provoked by geometric losses and attenuations while MISO systems are capable of mitigating the turbulence effects only. Consequently, It has been shown that by increasing the number of relays, the overall performance of the FSO system can be improved significantly.

\bibliographystyle{spmpsci}      
\bibliography{references}   
\end{document}